\documentclass[prd,aps,preprint,showpacs,superscriptaddress]{revtex4}
\usepackage{epsfig}
\usepackage{amsmath}

\begin{document}
\preprint{RBRC-671} \preprint{SLAC-PUB-12503}

\title{Effect of Orbital Angular Momentum on \\Valence-Quark Helicity Distributions}

\author{Harut Avakian}
\affiliation{Thomas Jefferson National Accelerator Facility, Newport News,
VA 23606}
\author{Stanley J. Brodsky}
\affiliation{Stanford Linear Accelerator Center, Stanford
University, Stanford, CA 94309}
\author{Alexandre Deur}
\affiliation{Thomas Jefferson National Accelerator Facility, Newport News,
VA 23606}
\author{Feng Yuan}
\affiliation{RIKEN/BNL Research Center, Building 510A, Brookhaven
National Laboratory, Upton, NY 11973}

\date{\today}
\vspace{0.5in}
\begin{abstract}
We study the quark helicity distributions at large $x$ in
perturbative QCD, taking into account contributions from the
valence Fock states of the nucleon which have nonzero orbital
angular momentum.  These states are necessary to have a nonzero
anomalous magnetic moment. We find that the quark orbital angular
momentum contributes a large logarithm to the negative helicity
quark distributions in addition to its power behavior, scaling as
$(1-x)^5\log^2(1-x)$ in the limit of $x\to 1$. Our analysis shows
that the ratio of the polarized over unpolarized down quark
distributions, $\Delta d/d$,  will still approach 1 in this limit.
By comparing with the experimental data, we find that this
ratio should cross zero at $x\approx 0.75$.
\end{abstract}

\maketitle

{\bf 1. Introduction.} Power-counting rules for the large-$x$ parton
distributions in hadrons have been derived more than three decades
ago based on perturbative quantum chromodynamics (pQCD) combined
with a $S$-wave quark model of hadrons
\cite{Gunion:1973nm,Blankenbecler:1974tm,Farrar:1975yb,{Lepage:1980fj}}.
The basic argument is that when the valence quark carries nearly
all of the longitudinal momentum of the hadron, the relevant QCD
configurations in the hadronic wave function become far off-shell
and can be treated in pQCD. A generic factorization has recently
been used to justify the power-counting rule by relating the
parton distributions at large-$x$ to the quark distribution
amplitudes of hadrons \cite{Ji:2004hz}. The power-counting rule
has also been generalized to sea quarks, gluons,
helicity-dependent distributions
\cite{Brodsky:1994kg,{Brodsky:2006hj}}, and generalized parton
distributions \cite{Yuan:2003fs}.

The leading pQCD diagrams associated with the leading Fock state
of the proton wave function predict that the positive helicity
(quark spin aligned with the proton spin) quark distribution
$q^+(x)$ scales as $(1-x)^3$, ($x=k^+/P^+$ is the light-cone
momentum fraction of the struck quark and is identical to the
Bjorken $x_B$ in the leading twist approximation), whereas the
negative helicity (quark spin anti-aligned with the proton spin)
quark distribution $q^-(x)$ is suppressed by $(1-x)^2$ relative to
the positive helicity one, scaling as $(1-x)^5$ at large $x$
\cite{Farrar:1975yb}. The direct consequence of these power laws
for the quark distributions is that the ratio of polarized quark
distribution $\Delta q(x)=q^+(x)-q^-(x)$ over the unpolarized
quark distribution $q(x)=q^+(x)+q^-(x)$ approaches 1 in the limit
$x\to 1$; i.e., at large $x$, $q^+$ dominates over $q^-$. When this
prediction is compared to the experimental data
\cite{Abe:1997dp,Zheng:2003un,{Dharmawardane:2006zd},{Airapetian:2004zf}},
it is interesting to observe that, for the up quark the ratio
increases with $x$, and seems to approach 1 at large $x$. However,
the ratio for the down quark is still far below 1, and remains
negative for a wide range of $x<0.6$ \cite{Zheng:2003un}. This
discrepancy has stimulated much theoretical interest.

In this paper we will reexamine the large-$x$ quark helicity
distributions in the perturbative QCD framework
\cite{Farrar:1975yb,{Lepage:1980fj}}. We work in light-cone gauge
with $A^+=0$, where there is no ghost contributions, and orbital
angular momentum is physical \cite{Brodsky:1997de}. We will take
into account the contributions from not only the leading
light-cone Fock state expansion of the nucleon wave function with
zero quark orbital angular momentum ($L_z=0$), but also the
valence Fock states with nonzero quark orbital angular momentum
($L_z\neq 0$). These contributions are naturally required to
obtain a nonzero anomalous magnetic moment for nucleons
\cite{Brodsky:1980zm} and are also present in the wave function
solutions in the AdS/CFT correspondence approach
\cite{Brodsky:2006uq}. We find that for the negative quark
helicity distribution $q^-$, there exist large logarithmic
enhancements from the $|L_z|=1$ Fock state component of the
proton. With this large logarithmic modification, we can explain
the discrepancy between the power-counting rule and experimental
data.

{\bf 2. Analysis of the large-$x$ behavior of the quark helicity
distributions.} In the $x \to 1$ regime where the struck quark has
nearly all of the light-cone momentum of its parent hadron, the
relevant QCD dynamics becomes far-off the mass shell: the Feynman
virtuality of the struck quark becomes highly space-like: $k^2_F -
m^2 \sim -{k^2_\perp + {\cal M}^2\over 1-x}$, where $k_\perp$ is
the transverse momentum of the struck quark and $\cal M$ is the
invariant mass of the spectator system. Thus we can use
perturbative QCD to analyze the large-$x$ behavior of the parton
distributions since the internal propagators in the relevant
Feynman diagrams scale as $1/(1-x)$. This behavior leads to the
power-counting rules. In fact, more partons in the hadron's wave
function mean more propagators in the scattering amplitudes and
more suppression for the contribution to the parton distributions.
Thus the parton distributions at large-$x$ depend on the number of
spectator partons in the Fock state wave function of the hadron.
For example, the valence quark distributions of nucleon will be
dominated by the three-quark Fock states of the nucleon wave
function. The three-quark Fock state expansion of the nucleon wave
function consists of zero orbital angular momentum component
($L_z=0$) and nonzero orbital angular momentum component ($L_z\neq
0$) \cite{Ji:2002xn}. In the following discussion, we will
consider the contributions from both Fock state components.

\begin{figure}[t]
\begin{center}
\epsfig{file=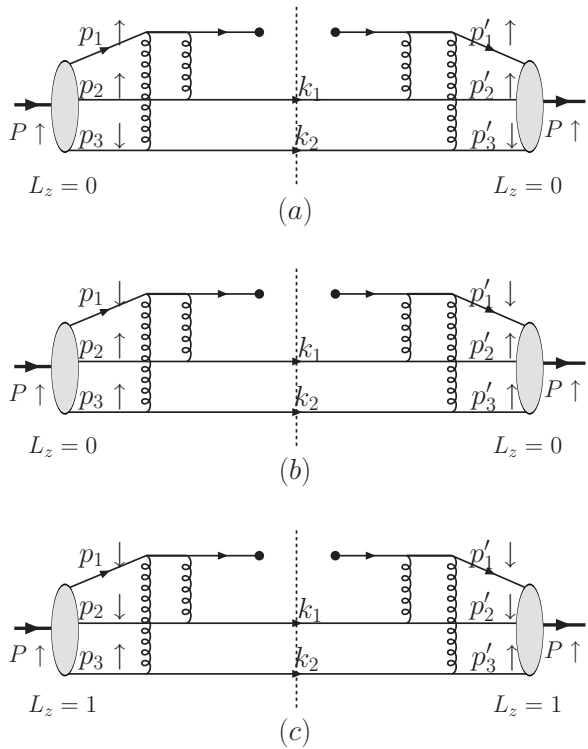,height=10.0cm,angle=0}
\end{center}
\caption{\it  Examples of Feynman diagrams which contribute to the
$q^\pm$ quark distributions at large $x$: (a) for $q^+$ with power
contribution of $(1-x)^3$; (b) for $q^-$ with $(1-x)^5$; (c) for
$q^-$ with $(1-x)^5\log ^2(1-x)$. The wave functions at the left
and right sides of the cut line in (a) and (b) represent the
leading Fock state expansion with zero quark orbital angular
momentum, whereas those for (c) represent the valence Fock state
with one-unit of quark orbital angular momentum.}
\end{figure}

In the original power-counting analysis of the quark helicity
distributions \cite{Brodsky:1994kg}, only the contributions from the leading Fock state
with $L_z=0$ have been taken into account. In Fig.~1(a,b) we show
the typical diagrams which contribute to the positive (a) and
negative (b) quark helicity distributions at large $x$. In terms of
the leading order quark distribution amplitude which corresponds
to the $L_z=0$ Fock state expansion of the proton wave function
\cite{Lepage:1980fj}, these quark helicity distributions can be
written as,
\begin{eqnarray}
q^\pm(x)|_{x\to 1}= \int[dy_i][dy_i'] \Phi(y_i)\Phi'(y_i'){\cal
H}^{(\pm)}\left(y_i,y_i';(1-x)\right) \ ,
\end{eqnarray}
where the integration measure $[dy_i]$ is defined as
$[dy_i]=dy_1dy_2dy_3\delta(1-y_1-y_2-y_3)$, and the $y_i$ are the
light-cone momentum fractions of the proton carried by the quarks
in the light-front wave functions, i.e., $p_i=y_iP$ and
$p_i'=y_i'P$ in Fig.~1. Here, $\Phi$ and $\Phi'$ represent the
quark distribution amplitudes of the proton at the left and right
sides of the cut line, respectively. ${\cal H}$ represents the
hard part of the amplitude which depends on $y_i$ and $y_i'$, and
$(1-x)$ as well.

The large-$x$ behavior of the hard factor ${\cal H}$ can be
evaluated from the partonic scattering amplitudes as shown by the
Feynman diagrams in Fig.~1. The wave functions corresponding to
Figs.~1(a,b) have zero quark orbital angular momentum, and thus
the total quark spin will be equal to the proton spin. If the
struck quark spin is the same as the proton spin (for the positive
helicity quark distribution $q^+$), the two spectator quarks will
be in the spin-0 configuration, whereas for the negative helicity
quark distribution $q^-$ with the quark spin opposite to the
proton spin, the two spectator quarks will be in the spin-1
configuration. It has been known for a long time that the hard
partonic part ${\cal H}$ with a spin-1 configuration for the two
spectators will be suppressed by $(1-x)^2$ relative to that with
spin-0 configuration
\cite{Farrar:1975yb,{Gunion:1983ay},{Brodsky:2006hj}}. That is
also the reason why the negative helicity quark distribution is
suppressed by $(1-x)^2$ relative to the positive helicity quark
distribution from this contribution.

The far-off-shell propagators in the partonic Feynman diagrams are
the dominant contributions to the power-counting of $(1-x)$ at
large $x$. One of the gluon propagators in Fig.~1 behaves as
\begin{equation}
\frac{1}{(p_3-k_2)^2}=\frac{1}{2p_3\cdot k_2}\approx
-\frac{1}{\langle k_\perp^2\rangle}\frac{1-x}{y_3} \ ,
\end{equation}
at large $x$. In the above expression, we have omitted all higher
order terms suppressed by $(1-x)$. Here, $\langle
k_\perp^2\rangle\sim \langle k_{1\perp}^2\rangle \sim \langle
k_{2\perp}^2\rangle$, represents a typical transverse momentum
scale for the spectator system. Each propagator will provide a
suppression factor of $(1-x)$. Counting the hard propagators in
Fig.~1(a), we find the total suppression factor from the hard
propagators is
\begin{equation}
\sim \frac{(1-x)^8}{y_2y_3(1-y_2)y_2'y_3'(1-y_2')} \ .\label{e23}
\end{equation}
We notice that the above expression does not introduce additional
dependence on $(1-x)$ upon integration over $y_i$ and $y_i'$,
assuming that the leading twist distribution amplitudes $\Phi$ and
$\Phi'$ are proportional to $y_1y_2y_3$ and $y_1'y_2'y_3'$
\cite{Lepage:1980fj}, respectively. Combining these results with
the power behavior for the other parts of the partonic scattering
amplitudes and the phase space integral, we find the positive
helicity quark distribution $q^+$ scales as $(1-x)^3$, whereas the
negative helicity quark distribution $q^-$ scales as $(1-x)^5$
\cite{Farrar:1975yb,{Lepage:1980fj},{Brodsky:1994kg},{Brodsky:2006hj}}.

In the above analysis, we only considered the contributions from
the leading Fock state of the proton with zero quark orbital
angular momentum. In general, the contributions from the higher
Fock states and the valence Fock states with nonzero quark orbital
angular momentum will introduce additional suppression in (1-x)
\cite{Lepage:1980fj,{Brodsky:2006hj}}. However, the
nonzero-quark-orbital-angular-momentum Fock state can provide
large logarithmic enhancement to the helicity flip amplitudes. For
example, it was found that the nonzero quark orbital angular
momentum contributes a large logarithmic term to the nucleon's
helicity-flip Pauli form factor $F_2(Q^2)$, which leads to the
scaling behavior $F_2(Q^2)\sim \log^2(Q^2/\Lambda^2)/Q^6$ at
$Q^2\to \infty$ \cite{{Ji:2003fw},Belitsky:2002kj}. This is
consistent with recent experimental data from JLab
\cite{Gayou:2001qd}. In the following, we will study the nonzero
quark orbital angular momentum contribution to the $q^-$ quark
distribution which is also associated with the helicity-flip
amplitude. The corresponding contributions to the positive quark
helicity distribution are always power suppressed
\cite{Brodsky:2006hj}.

In Fig.~1(c), we show an example of a contribution from the
$L_z=1$ Fock state of proton. Because  the quark orbital angular
momentum contributes one unit of the proton spin, we can have
difference between the total quark spin and the proton spin. If
the two spectator quarks are in the spin-0 configuration, this
will enhance the power-counting in the hard factor ${\cal H}$. On
the other hand, in order to get a nonzero contribution, we have to
perform the intrinsic transverse momentum expansion for the hard
partonic scattering amplitudes \cite{Belitsky:2002kj}, which will
introduce an additional suppression factor in $(1-x)$
\cite{Brodsky:2006hj}. For example, one of the contributions from
the diagram shown in Fig.~1(c) to the negative helicity quark
distribution will be proportional to
\begin{equation}
q^-(x)\propto \int (p_{1}^x+ip_1^y)(p_{1}^{\prime x}-ip_1^{\prime
y})\tilde{\psi}^{(3)}(y_i,p_{i\perp})\tilde{\psi}^{(3)}(y_i',p_{i\perp}')
T_H\left(y_i,p_{i\perp};y_i',p_{i\perp}'\right) \ ,
\end{equation}
where $\tilde{\psi}^{(3)}$ is a light-front wave function
amplitude for the $L_z=1$ Fock state of the proton
\cite{Ji:2002xn}. The intrinsic transverse momentum expansion is
essential to get the nonzero contributions. Otherwise, the
integral over the transverse momenta $p_{i\perp}$ and
$p_{i\perp}'$ will vanish because of the explicit factors
$p_{1}^x+ip_{1}^y$ and $p_{1}^{\prime x}+ip_{1}^{\prime y}$ in the
above equation. One intrinsic transverse momentum expansion comes
from the propagator we mentioned above,
\begin{eqnarray}
\frac{1}{(p_3-k_2)^2}&=&\frac{1}{(y_3P-k_2+p_{3\perp})^2}\nonumber\\
&\approx &\frac{\beta(1-x)}{y_3k_{2\perp}^2}\left(1-
\frac{\beta(1-x)}{y_3k_{2\perp}^2}2p_{3\perp}\cdot
k_{2\perp}\right) \ , \label{e24}
\end{eqnarray}
where $\beta$ is defined as $k_2^+/(1-x)P^+$, and we have kept the
linear dependence on $p_{3\perp}$ in the above expansion. Only
this linear term will contribute when integrating over
$p_{i\perp}$: $\int k_{2\perp}\cdot
p_{3\perp}(p_1^x+ip_1^y)\tilde{\psi}^{(3)}\propto (k_2^x+ik_2^y)
y_3\Phi_4(y_1,y_2,y_3)$, where $\Phi_4$ is one of the twist-4
quark distribution amplitudes of the proton
\cite{Belitsky:2002kj,{Braun:2000kw}}. From the above expansion,
we find that this term will introduce additional factor of
$(1-x)/y_3$ in the hard factor. Similarly, because of the
$p_1^{\prime x}-ip_1^{\prime y}$ factor in Eq.~(4), we have to do
the expansion in intrinsic transverse momentum associated with the
wave function at the right side of the cut line, and again the
expansion of the gluon propagator with momentum of $p_3'-k_2$ will
introduce another suppression factor of $(1-x)/y_3'$ in the hard
factor. Thus the total suppression factor from the above two
expansions will be $(1-x)^2/y_3y_3'$, which gives the same power
counting contribution to $q^-$ as that from the leading Fock state
with $L_z=0$ in the above.

We thus find the contributions from $L_z=1$ Fock state of the
proton do not change the power counting for the $q^-$ quark
distribution at large $x$. However, the additional factor
$1/y_3y_3'$ from the intrinsic transverse momentum expansions will
lead to a large logarithm when integrating over $y_i$ and $y_i'$.
This is because, combining the above two factors with all other
factors from the propagators shown in Eq.~(\ref{e23}), the total
dependence on $y_i$ and $y_i'$ for the hard factor will be
\begin{equation}
\sim\frac{1}{y_2y_3^2(1-y_2)y_2'y_3^{\prime 2}(1-y_2')} \ ,
\end{equation}
where we have $y_3^2$ and $y_3^{\prime 2}$ in the denominator. On
the other hand, we expect the twist-4 quark distribution
amplitude to have the following behavior at the end point region:
$y_3\Phi_4(y_1,y_2,y_3)\propto y_1y_2y_3$ and
$y_3'\Phi_4(y_1',y_2',y_3')\propto y_1'y_2'y_3'$
\cite{Braun:2000kw}. Thus we will have logarithmic divergences for
the integrations over $y_3$ and $y_3'$, for which we can
regularize in terms of $\log(1-x)$ as indicated in the above
propagator expansion. This will lead to a double logarithmic
contribution $\log^2(1-x)$ in addition to the power term
$(1-x)^5$ to the $q^-$ quark distribution at large $x$.

In summary, for the negative quark helicity distribution $q^-$ at
large $x$, the leading Fock state with zero quark orbital angular
momentum $L_z=0$ contributes to a power term  $(1-x)^5$, whereas
the valence Fock state with $|L_z|=1$ contributes to a double
logarithmically enhanced term $(1-x)^5\log^2(1-x)$. So, in the
limit $x\to 1$, the $q^-$ distribution will be dominated by the
contributions from $L_z=1$ Fock state of the proton, scaling as
$(1-x)^5\log^2(1-x)$. In the intermediate x range, the sub-leading
terms can also be important. For example in
Ref.~\cite{Brodsky:1994kg}, the quark helicity distributions were
parameterized by the leading and sub-leading power terms and fit
to the experimental data. This was later updated to account for
the latest data in Ref.~\cite{Leader:1997kw}. Thus, as a first
step towards a comprehensive phenomenology, we follow the
parameterizations for $q^+$ and $q^-$ in
Ref.~\cite{Brodsky:1994kg} by adding the newly discovered double
logarithms enhanced contributions,
\begin{eqnarray}
u^+(x)&=&\frac{1}{x^\alpha}\left[A_u(1-x)^3+B_u(1-x)^4\right]
\nonumber\\
d^+(x)&=&\frac{1}{x^\alpha}\left[A_d(1-x)^3+B_d(1-x)^4\right]
\nonumber\\
u^-(x)&=&\frac{1}{x^\alpha}\left[C_u(1-x)^5+C_u'(1-x)^5\log^2(1-x)+D_u(1-x)^6\right]
\nonumber\\
d^-(x)&=&\frac{1}{x^\alpha}\left[C_d(1-x)^5+C_d'(1-x)^5\log^2(1-x)+D_d(1-x)^6\right]
\ , \label{e25}
\end{eqnarray}
where the additional two parameters $C_u'$ and $C_d'$ come from
the logarithmic modifications to the $q^-$ quark distribution at
large $x$, and all other parameters refer to \cite{Brodsky:1994kg}.
In the following, we will fit to the current experimental data at
large $x$ region with the above parameterizations for the valence up
and down quarks.

{\bf 3. Phenomenological applications.} In order to demonstrate
the importance of the new scaling behavior for the negative
helicity distributions for the valence up and down quarks, we
analyze the latest experimental data from SLAC, HERMES and
Jefferson Lab, including Hall A and Hall B data
\cite{Abe:1997dp,Zheng:2003un,{Dharmawardane:2006zd},{Airapetian:2004zf}}.
We will keep the original fit values for other parameters
\cite{Leader:1997kw} except the two new parameters: $C_u'$ and
$C_d'$. We only use the experimental data in the large-$x$ region,
i.e., $x>0.3$, where the sea contribution is not significant. We
perform our fit at a fixed $Q^2=4$ GeV$^2$, and all the
experimental data are evolved to this scale by using the GRSV
parameterization \cite{Gluck:2000dy} for the polarized and
unpolarized quark distributions. The evolution introduces some
theoretical uncertainties.

\begin{figure}[t]
\begin{tabular}{cc}
\includegraphics[height=6.0cm,angle=0]{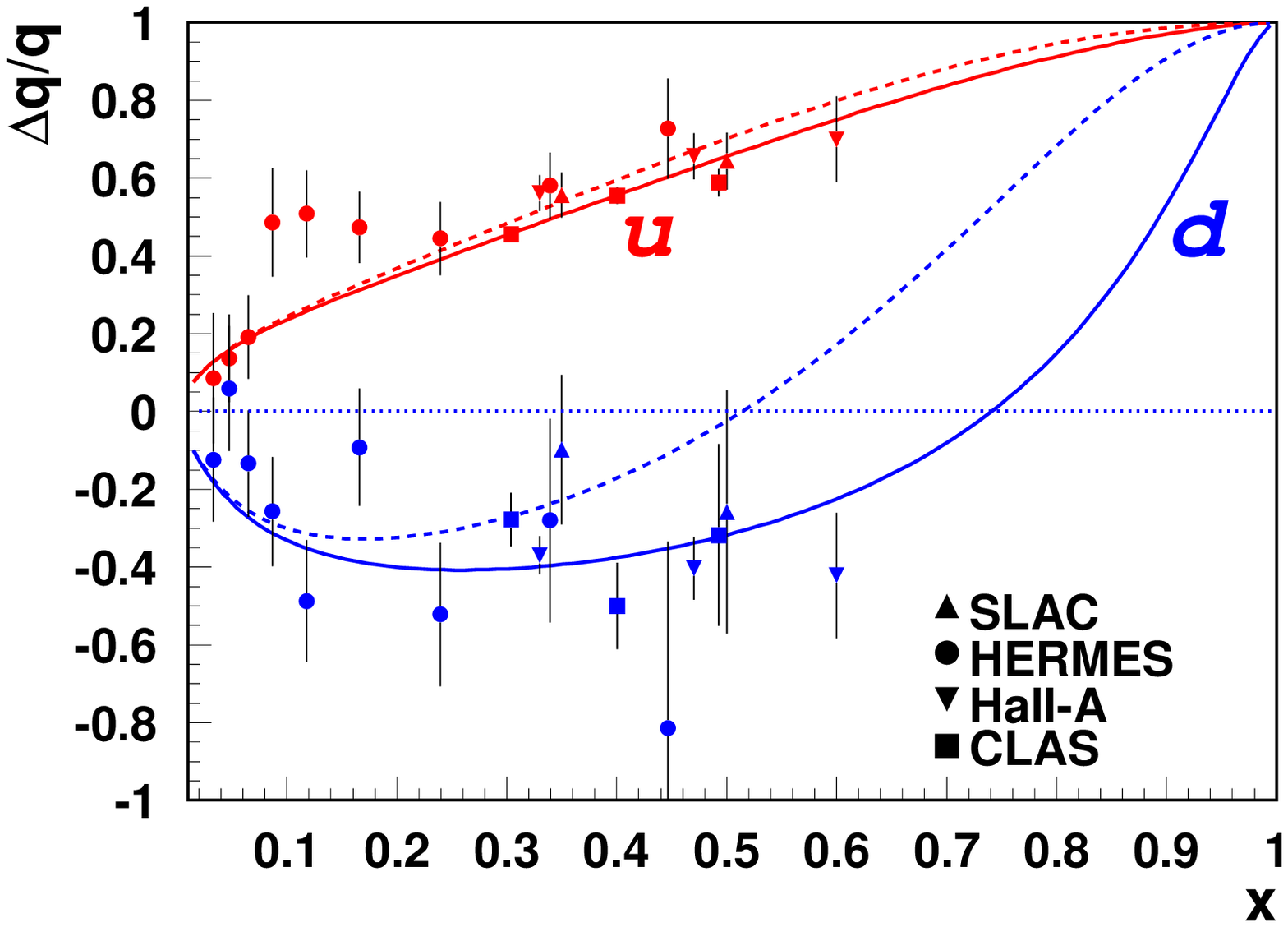}
&
\includegraphics[height=6.0cm,angle=0]{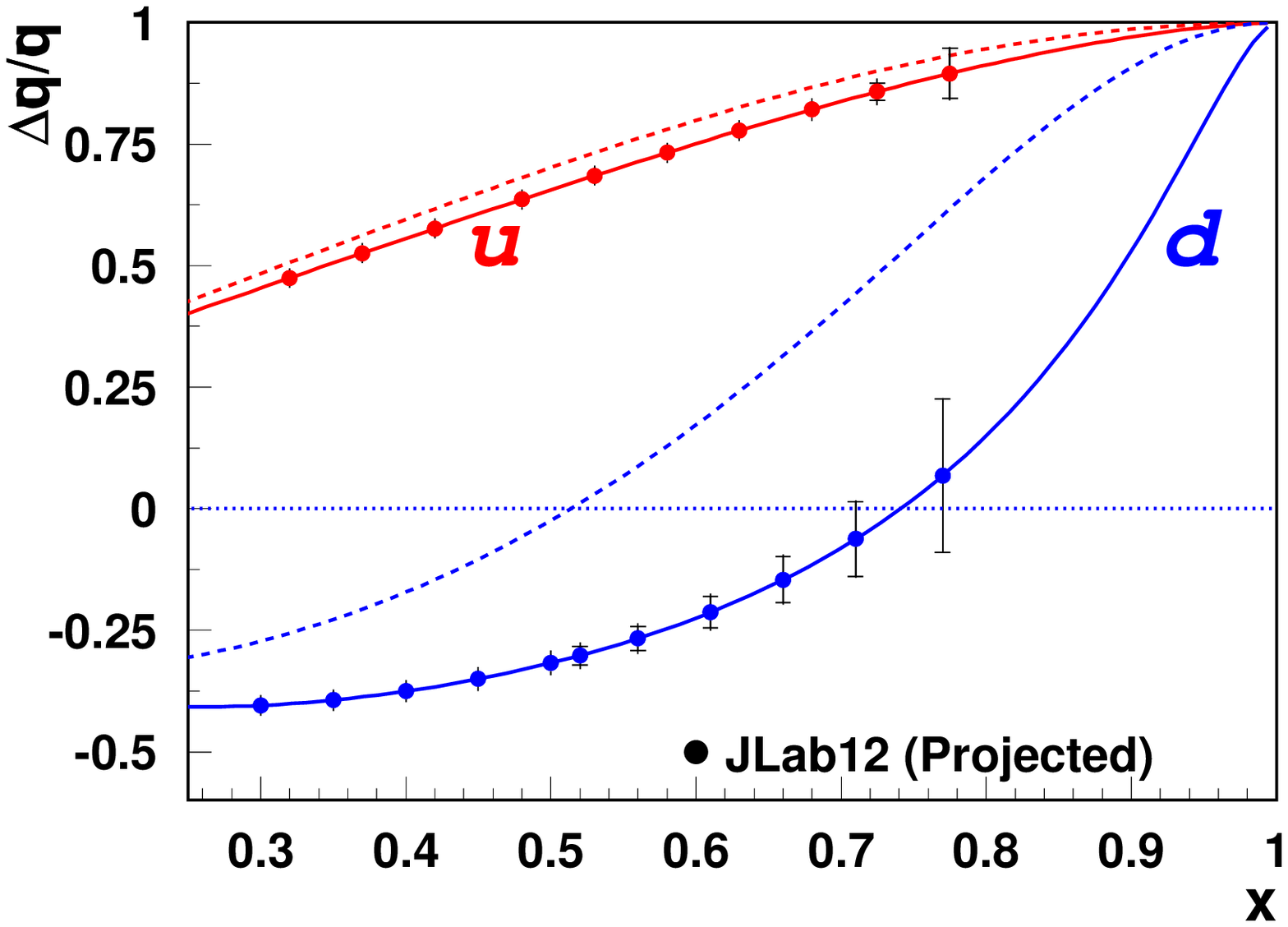}
\end{tabular}
\caption{\it Comparison of the quark helicity distributions
Eq.~(\ref{e25}) with experimental data (left panel), and future
projections from JLab (right panel) as functions of $x$ for up
(the upper curves) and down (the lower curves) quarks. The circles
are for HERMES data \cite{Airapetian:2004zf}, the triangles  up
for SLAC \cite{Abe:1997dp}, the triangles down for JLab Hall-A
data \cite{Zheng:2003un}, the filled squares for CLAS
\cite{Dharmawardane:2006zd}, and open squares for 12 GeV upgrade
projection for CLAS. The dashed curves are the predictions from
\cite{Leader:1997kw}, and the solid ones are our fit results (only
the large-$x$ ($>0.3$) experimental data were used in the fit). The
symbols in the right panel show combined projections from all
three JLab experiments \cite{PAC32}. }
 \label{fig:proj}
\end{figure}

From our fit, we find the following values for $C_u'$ and $C_d'$,
\begin{equation}
C_u'=0.493 \pm 0.249,~~~~C_d'=1.592\pm0.378 \ , \label{e26}
\end{equation}
which are comparable in size to $C_u=2.143\pm0.137$ and
$C_d=1.689\pm0.227$ in Ref.~\cite{Leader:1997kw}. The minimum of
the functional $\chi^2$ is achieved at $\chi^2=11.4$ and
$\chi^2/DOF=11.4/10=1.14$. We further notice that the additional
two terms in Eq.~(\ref{e25}) do not change significantly the sum
rules for the up and down quarks, such as the Bjorken and momentum
sum rule, which are essential for constraining the parameters in
Refs.~\cite{Brodsky:1994kg,{Leader:1997kw}}. For example, they
contribute $\sim 4\%$ to the momentum sum rule coming from the
quarks.

In the left panel of Fig.~\ref{fig:proj}, we show the above fit,
where we plot the ratio of the polarized quark distribution
$\Delta q$ over the unpolarized quark distributions $q$ as
functions of $x$ for both up and down quarks, compared with the
experimental data. From these comparisons, we find that the ratio
for the up quark $\Delta u/u$ can still be described by the
parameterization based on the original power counting rule for
$u^+$ and $u^-$. This can also be seen from the small value of
$C_u'$ in our fit Eq.~(\ref{e26}), with big error bar though.
However, for the down quark we have to take into account a large
contribution from the newly discovered term for the negative
helicity distribution $d^-$; the difference between our result and
the original power-counting-rule inspired parameterization
\cite{Leader:1997kw} becomes significant at $x\gtrsim 0.5$. The
analysis of the anomalous magnetic moment and generalized parton
distributions of nucleons also indicates significant contributions
from the orbital angular momenta of up and down
quarks~\cite{Burkardt:2005km}. This is in qualitative agreement
with our fitting results, taking into account the large error bar
for $C_u'$. A precision determination of these contributions shall
be obtained by further development for a consistent set of
parameters for Eq.~(\ref{e25}) from next-to-leading-order QCD
analysis of both polarized and unpolarized data over the full
range in $x$ \cite{Leader:1997kw}.

Another important prediction of our fit is that the ratio of
$\Delta d/d$ will approach 1 at extremely large $x$, and it will
cross zero at $x\approx 0.75$. It will be interesting to check
this prediction in future experiments, such as the 12~GeV upgrade
of Jefferson Lab. For comparison, in the right panel of Fig.~2, we
show the experimental projections for these measurements from the
12 GeV upgrade of JLab experiments~\cite{PAC32}, together with our
predictions and the results from the previous power-counting-rule
parameterizations~\cite{Leader:1997kw}.

We thank N.~Akopov, P.~Bosted, J.P.~Chen, V.~Dharmawardane,
Z.-D.~Meziani and X.~Zheng for useful conversations on the
experimental data and many related discussions. We also thank
X.~Ji and W.~Vogelsang for their comments. This work was supported
by by the  United States Department of Energy. Jefferson Science
Associates (JSA) operates the Thomas Jefferson National
Accelerator Facility for the U. S. DOE under contract
DE-AC05-060R23177. F.Y. is grateful to RIKEN, Brookhaven National
Laboratory and the U.S. DOE (grant number DE-FG02-87ER40371 and
contract number DE-AC02-98CH10886) for providing the facilities
essential for the completion of this contribution.


\begin{thebibliography}
\frenchspacing

\bibitem{Gunion:1973nm}
J.~F.~Gunion,
Phys.\ Rev.\ D {\bf 10}, 242 (1974).

\bibitem{Blankenbecler:1974tm}
R.~Blankenbecler and S.~J.~Brodsky,
Phys.\ Rev.\ D {\bf 10}, 2973 (1974).

\bibitem{Farrar:1975yb}
G.~R.~Farrar and D.~R.~Jackson,
Phys.\ Rev.\ Lett.\  {\bf 35}, 1416 (1975).

\bibitem{Lepage:1980fj}
G.~P.~Lepage and S.~J.~Brodsky,
Phys.\ Rev.\ D {\bf 22}, 2157 (1980).

\bibitem{Ji:2004hz}
  X.~Ji, J.~P.~Ma and F.~Yuan,
  Phys.\ Lett.\  B {\bf 610}, 247 (2005).


\bibitem{Brodsky:1994kg}
S.~J.~Brodsky, M.~Burkardt and I.~Schmidt,
Nucl.\ Phys.\ B {\bf 441}, 197 (1995).

\bibitem{Brodsky:2006hj}
  S.~J.~Brodsky and F.~Yuan,
  Phys.\ Rev.\  D {\bf 74}, 094018 (2006).


\bibitem{Yuan:2003fs}
F.~Yuan,
Phys.\ Rev.\ D {\bf 69}, 051501 (2004).


\bibitem{Zheng:2003un}
  X.~Zheng {\it et al.},
  Phys.\ Rev.\ Lett.\  {\bf 92}, 012004 (2004);
  Phys.\ Rev.\  C {\bf 70}, 065207 (2004).

\bibitem{Dharmawardane:2006zd}
  K.V.~Dharmawardane {\it et al.},
  Phys.\ Lett.\,  {\bf  B641},11 (2006).

\bibitem{Airapetian:2004zf}
  A.~Airapetian {\it et al.},
  Phys.\ Rev.\  D {\bf 71}, 012003 (2005).
\bibitem{Abe:1997dp}
  K.~Abe {\it et al.}  [E154 Collaboration],
  Phys.\ Lett.\  B {\bf 405}, 180 (1997).
\bibitem{Brodsky:1997de}
  S.~J.~Brodsky, H.~C.~Pauli and S.~S.~Pinsky,
  Phys.\ Rept.\  {\bf 301}, 299 (1998).

\bibitem{Brodsky:1980zm}
  S.~J.~Brodsky and S.~D.~Drell,
  Phys.\ Rev.\  D {\bf 22}, 2236 (1980).

\bibitem{Brodsky:2006uq}
  S.~J.~Brodsky and G.~F.~de Teramond,
  Phys.\ Rev.\ Lett.\  {\bf 96}, 201601 (2006).

\bibitem{Ji:2002xn}
  X.~Ji, J.~P.~Ma and F.~Yuan,
  Nucl.\ Phys.\  B {\bf 652}, 383 (2003).

\bibitem{Gunion:1983ay}
  J.~F.~Gunion, P.~Nason and R.~Blankenbecler,
  Phys.\ Rev.\  D {\bf 29}, 2491 (1984).

\bibitem{Ji:2003fw}
  X.~Ji, J.~P.~Ma and F.~Yuan,
  Phys.\ Rev.\ Lett.\  {\bf 90}, 241601 (2003).


\bibitem{Belitsky:2002kj}
  A.~V.~Belitsky, X.~Ji and F.~Yuan,
  Phys.\ Rev.\ Lett.\  {\bf 91}, 092003 (2003).

\bibitem{Gayou:2001qd}
  O.~Gayou {\it et al.},
  Phys.\ Rev.\ Lett.\  {\bf 88}, 092301 (2002).


\bibitem{Braun:2000kw}
  V.~Braun, R.~J.~Fries, N.~Mahnke and E.~Stein,
  Nucl.\ Phys.\  B {\bf 589}, 381 (2000).

\bibitem{Leader:1997kw}
  E.~Leader, A.~V.~Sidorov and D.~B.~Stamenov,
  Int.\ J.\ Mod.\ Phys.\  A {\bf 13}, 5573 (1998).

\bibitem{Gluck:2000dy}
  M.~Gluck, E.~Reya, M.~Stratmann and W.~Vogelsang,
  Phys.\ Rev.\  D {\bf 63}, 094005 (2001).

\bibitem{PAC32}
Jefferson Lab Hall A, B. Wojtsekhowski et~al.,
\newblock JLab Proposal E12-06-122 (2006); Hall B, . S.Kuhn et~al.,
\newblock E12-06-109 (2006); Hall C, X.Zheng et~al.,
\newblock E12-06-110 (2006).

\bibitem{Burkardt:2005km}
  M.~Burkardt and G.~Schnell,
  Phys.\ Rev.\  D {\bf 74}, 013002 (2006).

\end{thebibliography}

\end{document}